\documentstyle[12pt]{article}
\if@twoside
 \else
 \fi
 
 \parskip \medskipamount
\begin{document}
\hbox{}
\nopagebreak
\vspace{-3cm}
\addtolength{\baselineskip}{.8mm}
\baselineskip=24pt
\begin{flushright}
{\sc UMN-TH-1513-96}\\
{\sc TPI-MINN-96/16-T}\\
hep-ph/9612343\\
September  1996
\end{flushright}

\begin{center}
{\Large  Dual superconductivity. Variations on a theme.}\\
\vspace{0.1in}

\vspace{0.5in}
{\large Alex Kovner}\\
{\it Physics Department, University of Minnesota\\
116 Church st. S.E., Minneapolis, MN 55455, USA}\\
\vspace{0.1in}

\vspace{1in}

{\sc  Abstract} 
\end{center}

\noindent
It is pointed out that the low energy effective 
theory that describes the low lying glueballs
of the pure Yang Mills theory sustains static classical stringlike solutions.
We suggest that these objects can be identified with the QCD flux tubes and 
their energy per unit
length with the string tension.

\vfill

\newpage

It is a great pleasure for me to write this paper in honor of Larry Horowitz. 
Larry has a rare gift
of being able to inspire 
and encourage people around him even when their work has little
overlap with his main 
line of research. I for one, was very lucky to experience the invigorating
atmosphere of long and frequent discussions with him, 
and much of my work is directly or indirectly influenced by it.
One of the many research 
areas of Larry's is the study of topological objects in field theory, 
and in particular 
in QCD. This perhaps is where my own research is closest to his.
I have decided therefore 
to devote this paper to a subject which belongs to this area - the dual
superconductor model of confinement in QCD.

Confinement is widely believed to be a property of strong interactions. 
It is certainly one of the most
striking properties of the strong interactions. It is therefore somewhat 
ironic that it has 
also proved to be the most difficult one to understand theoretically.  
So much so that it is 
not even clear what objects exactly are confined by strong interactions.

 A seemingly straightforward answer to this question is that those are 
perturbative quarks and 
gluons. A little reflection, however shows that this answer is totally 
unsatisfactory. 
Quarks and gluons are the main objects of perturbative QCD, and as such 
are very helpful 
concepts in the perturbative context. However in the strict sense quarks 
and gluons can not
even in principle exist as asymptotic states. This is due to the fact that 
they carry 
nontrivial color charges. The color group is not a bona fide physical symmetry
 of the
strong interactions, but is a gauge symmetry. Which means that it is not a 
symmetry of physical
states but rather a 
redundancy of our way of description of strong interaction physics in terms 
of the standard QCD 
Lagrangian. Physical states are by construction color singlets and therefore
can not contain a single 
quark or a gluon. As opposed to abelian gauge theories, where the global
part of the gauge 
group is indeed a symmetry of physical states, in nonabelian theories not 
even the global 
color transformation is physical. So perturbative partons are confined by 
construction, and 
this confinement does not seem to depend on the dynamics of the theory. 
For example, to the best of my understanding this ``kinematical''
 confinement will 
be also present in the $SU(3)$ theory with more than 16 flavors, which is 
infrared free perturbatively, 
and which is not expected to be confining in the usual colloquial
sense. What we mean by confinement must be some other genuinely
dynamical effect.

Even though 
in non Abelian theories confinement is a somewhat undefined concept, 
it is
intuitively clear that there is a dynamical phenomenon to be understood
here.
Many models of confinement have been put forth.
Perhaps the most widely accepted one is the dual superconductor model due to
t'Hooft and Mandelstam \cite{dual}. 
In this note I want to discuss one particular
objection to this model if understood literally, and suggest a related
but in certain aspects very different model which also exhibits the 
property of confinement and is motivated by the low energy spectrum of a pure
Yang Mills theory. 

Let me first briefly review the idea of dual superconductor.
The idea itself is very simple and as such certainly very appealing.
Consider a superconductor. It is described by the Landau - Ginzburg model
of the complex order parameter field $H$ (the Higgs field) coupled to
vector potential $A_i$. 
The free energy, or Lagrangian of this Abelian Higgs model (AHM)
in the field
theoretical context, for static field configurations is
\begin{equation}
-L=|D_iH|^2+{1\over 2}B^2+\lambda (H^*H-v^2)^2
\end{equation}
where
\begin{equation}
D_i=\partial_i-igA_i
\end{equation}
The superconducting phase is characterized by a nonvanishing condensate of the
Higgs field
\begin{equation}
<H>=v
\end{equation}
In this phase it is convenient to work in the unitary gauge, that is to rewrite
Lagrangian in terms of the field 
\begin{equation}
a_i=A_i-{1\over g}\partial_i\phi
\end{equation}
where
\begin{equation}
H=\rho\exp\{i\phi\}
\label{phi}
\end{equation}
The Lagrangian then becomes
\begin{equation}
-L=(\partial_i\rho)^2+g^2\rho^2a_i^2+{1\over 2}B^2+\lambda (\rho^2-v^2)^2
\end{equation}

As is well known, the only way the magnetic field can penetrate a
superconductor is in the form of the flux tubes with the core size of the
penetration depth $l=(gv)^{-1}$.
We will recap the derivation of this property here, even though it is
very well known since we will need to refer to some of its features later.

Let us consider a superconductor of the linear dimension L, and try to find
the configuration with minimnal energy subject to condition that it sustains
magnetic flux $\Phi$ in the direction of the z axis.
We also assume of course that $L>>l$.
In the absence of the condensate the vector field is massless. The magnetic 
field inside the medium therefore is basically the field of a dipole of the
size $L$,
and for $r_{perp}=\sqrt{x^2+y^2}<<L$ behaves as 
$B_z\propto{ r_{perp}^2\over L^4}$. The 
vector potential is $a_i\propto\epsilon_{ij}r_jr^2/L^4$, 
and the energy of such a configuration is finite
in the limit of large $L$. Now consider the superconducting state. The vector
field is massive. 
The quadratic part of the Lagrangian becomes
\begin{equation}
-L=(\partial_i\sigma)^2+g^2v^2a_i^2+{1\over 2}B^2+m^2\sigma^2
\label{super}
\end{equation}
with $\sigma=\rho-v$.
The vector field $a$ therefore can not possibly have such 
long range form in a 
state with minimal energy as in the nonsuperconducting state.
Instead, it should vanish 
exponentially as $\exp\{-gvr_{perp}\}$ for $r_{perp}>l$.
At first sight
this seems impossible, since a vector potential can vanish so 
fast only in a state with vanishing magnetic flux.
\begin{equation}
\Phi_z=\int_S d^2r_{perp}\epsilon_{ij}\partial_i a_j=\oint_{C=dS}dr_i
\epsilon_{ij}a_j=0
\end{equation}
Consider however a vector 
potential which far enough from the origin ($r_{perp}=0$) 
vanishes everywhere except on a half plain (let's say
$x=0, y>0$):
\begin{equation}
a_i=\Phi_z\delta_{1i}\delta(x)\theta(y)
\end{equation}
It certainly describes 
a finite magnetic flux $\Phi_z$. The singularity in $a_i$
also does not contribute to the energy, if the singularity is quantized. The
reason for that bizarre sounding 
statement is that the vector field mass term
is strictly speaking not $g^2v^2a_i^2$ as 
it is written in eq.\ref{super}. Remember that
it really is the remnant of the covariant derivative term $|D_iH|^2$. This
expression when written in terms of the phase $\phi$ is 
$g^2\rho^2(A_i-{1\over g^2}\partial_i\phi_{mod 2\pi})^2$. It does not
feel jumps in the phase $\phi$ which are integer multiples of $2\pi$. Since our
field $a_i$ was defined shifting $A_i$ by $\partial_i\phi$ the mass term in the
Lagrangian eq.\ref{super} is actually
\begin{equation}
g^2v^2(a^i_{mod 2\pi\Delta})^2
\end{equation}
where $\Delta$ is the ultraviolet regulator - lattice spacing.
For smooth fields $a_i$ there is no difference between the two mass terms. 
However eq.() admits singularities of the type of eq.(9).
The singularity of this type therefore does not cost energy provided 
\begin{equation}
\Phi_z={2\pi n\over g}
\end{equation}
with integer $n$.
For these values of the 
magnetic flux the lowest energy configuration has the form
$a_i=\Phi_z\delta_{1i}\delta(x)\theta(y)+b_i(x)$ 
with $b_i$ a smooth, exponentially decreasing
function. It represents a tube of magnetic flux along the $z$ axis, or 
Abrikosov-Nielsen-Olesen (ANO) vortex.
Properties of this solution are discussed, for example in \cite{vortex}.

So far we were discussing an ordinary superconductor.
Suppose now, that in addition to the charged field $\Phi$ our model also has
very heavy magnetic monopoles. Since they are very heavy, they do
not affect the dynamics of the order parameter. But the structure of the vacuum
does affect strongly the interaction between the monopoles themselves. A
monopole and an antimonopole in the superconducting vacuum feel linear
confining potential. Since a monopole is a source of magnetic flux, this 
flux in the superconducting 
vacuum will form a ANO flux tube, which terminates on 
an antimonopole. The energy of the flux tube is proportional to its
length, hence the linear potential.

So, here is a theory in which certain objects are confined by a linear 
potential. 
The dual superconductor hypothesis asserts 
that the Yang Mills theory is basically an ``upside down'' version
of the Abelian Higgs model, as far as the confinement mechanism is concerned.
That is, if we use the following dictionary, the preceding discussion
describes confinement in any non Abelian gauge theory, which is in the 
confining phase.
The magnetic field of the Higgs model should be called the `''color electric 
field'', the magnetic monopoles should be called the ``color charges'', and by
analogy the electrically charged field $H$ should turn into
``color magnetic monopole''. 

There are of course many questions that can be immediatelly asked.
The first thing is who are those mysterious monopoles, which do not even appear
in the QCD Lagrangian. 
But perhaps those could appear as a result of a duality transformation. One can
certainly draw some encouragement from the simplest known confining theory - 
Abelian compact $U(1)$ model.
This theory on the lattice is known to have a phase transition at some finite
value of the coupling constant. In the strongly coupled phase it is confining.
Note that in the Abelian, as opposed to non Abelian case, the global part
of the gauge group is a physical symmetry. Charged states do exist in the 
Hilbert space a priori, and confinement  is a purely dynamical effect.

Now, confinement in this theory does have a direct interpretation in
terms of dual superconductivity. In fact, in the lattice theory one can
perform a duality transformation which transforms the compact $U(1)$
pure gauge theory into a noncompact $U(1)$ theory with charged fields, i.e.
Abelian Higgs model \cite{u1}. 
The ``Higgs'' field in this model actually describes
excitations carrying magnetic charge - magnetic monopoles. Those are condensed
in the confining phase and dual superconductivity captures perfectly the 
physics of confinement in this theory.

The example of the $U(1)$ theory encouraging as it is, still leaves many 
questions unanswered. In non Abelian theories there is no gauge invariant
definition of a monopole, or for that matter even of the flux that it carries.
Since even the definition 
of monopoles depends on the gauge, it is unavoidable that the correlation
between the quantitative properties 
of these monopoles (e.g. monopole density) and confinement
(the value of the string tension) will also vary from gauge to gauge.
This is indeed known to 
be the case. In the so called maximal Abelian gauge these correlations are
indeed very strong\cite{maximal}. This gauge is designed to make the
gauge fixed Yang Mills 
Lagrangian to be as similar as possible to the compact $U(1)$ theory by 
attempting to maximally 
supress the fluctuations of all fields except the colour components 
of the gauge potential which belong to the Cartan subalgebra. In 
particular it seems that the monopoles contribute a large share of the string 
tension \cite{maximal}. Monopole
condensate disappears above the 
critical temperature in the deconfined phase. An interesting recent
work attempts to calculate  
the ``effective potential'' for the monopole field, and the results are
consistent with 
the ``Mexican hat'' type potential below the critical temperature and a single
well above $T_c$ \cite{orderp}. 

In other gauges 
correlations between the monopole properties and the confining properties of 
the theory are much 
weaker, and in some cases are completely absent\cite{various}.
Even in the maximal 
Abelian gauge it is not 
clear that the monopole dominance is not a lattice
artifact. One expects 
that if the monopoles are indeed physical objects in the continuum limit,
their size would be of order 
of the only dimensional parameter in the theory $\Lambda_{QCD}^{-1}$.
The monopoles which 
appear in the context of lattice calculations on the other hand are all
pointlike objects.

These are worrysome questions, 
but unfortunately I do not have anything new to add to these points.
The point that will 
concern me in this note is a different one. One can formulate yet another
objection to the dual 
superconductor picture. The dual superconductor hypothesis in effect states
that the effective low energy theory of the pure Yang Mills theory
when written in terms 
of appropriate field variables is an Abelian Higgs model. 
If so, the low lying 
spectrum of pure Yang Mills should be the same as that of an Abelian 
Higgs model.

A statement about the 
form of the effective Lagrangian is not easily verified, since it 
necessitates an appropriate 
choice of variables. On the other hand a statement about the spectrum
does not depend on the 
choice of variables and is in this sense universal and easily verifiable.
The spectrum of the 
Abelian Higgs model is well known. The two lowest mass excitations are the
massive vector particle and a massive scalar particle. In fact, the 
role of the vector particle was 
crucial in the discussion of confinement in the beginning of this paper.
The spectrum of the 
Yang Mills theory is not known from analytical calculations. However, in 
recent years a rather clear picture of it emerged from the lattice QCD 
simulations \cite{glueballs}.
The lowest lying particle in 
the spectrum is a scalar glueball with the mass $1.5-1.7$ Gev.
The second excitation 
is a spin 2 tensor glueball with a mass of around $2.2$ Gev.
Vector (and pseudovector) 
glueballs are conspicuously missing in the lowest lying part
of the spectrum. 
The simulations indicate that they are relatively heavy with masses above
2.8 Gev \cite{glueballs}. The pattern of 
the Yang Mills spectrum therefore seems to be rather different than the
one suggested by the Abelian 
Higgs model. Moreover, since the vector glueball is so much heavier
than the scalar and 
tensor ones, it seems very unlikely that it plays so prominent a role
in the confinement 
mechanism as the one played by the massive photon in a superconductor.

This of course can 
not be completely ruled out. It could happen that for some exotic reason,
it is the high 
energy part of the spectrum that has closer relation to confinement than
the lowest lying excitations.
However it seems more 
reasonable to assume that the low energy spectrum of the YM 
theory should serve as a guide
to the confinement mechanism.  
In that case it is almost inevitable that the spin 2 glueball 
should play a very important role. 
In fact, it is possible 
that the role of the spin 1 massive photon of the Abelian Higgs model
is played by the spin 2 glueball in the real life YM.
The purpose of this 
note is precisely to present this possibility and to explore it on a very
simplistic level. 
In what follows we will write down an effective theory, which should 
on very general grounds 
describe the dynamics of the spin 0 and spin 2 glueballs. We will then
show that in this theory 
certain external objects that carry an analog of the magnetic flux
are indeed confined. 
The tensor field that describes the spin 2 glueball causes this confinement
in a way very similar to 
the massive photon field in AHM. We will also show that this model
has other  similarities 
with AHM, so that even though its field content is very different,
in a certain sense it can be considered as a variant or extension
of a dual superconductor model.

We start therefore 
by writing down a theory which contains a scalar field $\sigma$ and a massive
symmetric tensor field $G_{\mu\nu}=G_{\nu\mu}$ with a simple interaction. 
\begin{equation}
L={1\over 4}G_{\lambda\sigma}
D^{\lambda\sigma\rho\omega}G_{\rho\omega}+\partial_\mu
\sigma\partial^\mu\sigma-2g^2 v\sigma G^{\mu\nu}G_{\mu\nu}
-g^2\sigma^2  G^{\mu\nu}G_{\mu\nu}-V(\sigma)
\label{model}
\end{equation}

The operator $D$ which appears in the kinetic term of the tensor field is
\begin{eqnarray}
D^{\lambda\sigma\rho\omega}&=&(g^{\lambda\rho}g^{\sigma\omega}+
g^{\sigma\rho}g^{\lambda\omega})(\partial^2+m^2)
-2g^{\lambda\sigma}g^{\rho\omega}(\partial^2+M^2) \\
&-&(\partial^\lambda\partial^\rho 
g^{\sigma\omega}+\partial^\sigma\partial^\rho g^{\lambda\omega}+
\partial^\lambda\partial^\omega 
g^{\sigma\rho}+\partial^\sigma\partial^\omega g^{\lambda\rho})
+2(\partial^\lambda\partial^\sigma 
g^{\rho\omega}+\partial^\omega\partial^\rho g^{\sigma\lambda})
\nonumber
\label{D}
\end{eqnarray}

Several comments 
are in order here. A general symmetric tensor field  has ten components.
A massless spin 
two particle has only two degrees of freedom. The tensor structure of the
kinetic term for the massless
tensor particles 
is therefore determined so that it should project out two components out of
$G^{\mu\nu}$. In fact it is easy to check that the kinetic term in eq.(\ref{D})
in the massless case 
($m^2=M^2=0$) is invariant under the four parameter local gauge transformation
\begin{equation}
\delta G^{\mu\nu}=\partial_\mu\Lambda_\nu+\partial_\nu\Lambda_\mu
\label{gauge}
\end{equation}
These four gauge invariances together with four
corresponding gauge fixing conditions indeed eliminate
eight components 
of $G^{\mu\nu}$. This gauge invariance is broken by the mass. However the
equations of motion even in the massive case lead to four constraints
\begin{equation}
\partial_\mu G^{\mu\nu}=0
\label{constr}
\end{equation}

This eliminates four 
degrees of freedom out of ten\footnote{In the interacting theory, 
eq.\ref{model}, the 
constraint equations are slightly modified, but they still provide four
conditions on the fields.}. In addition the scalar field $G^\mu_\mu$
decouples from the rest 
of the dynamics and can be neglected. In fact we have added a parameter 
$M$ whose only purpose 
is to make $G^\mu_\mu$ arbitrarily heavy. All in all therefore we are left
with five independent 
propagating degrees of freedom in $G^{\mu\nu}$, which is the correct number
to describe a massive spin two particle.

The scalar 
glueball self interaction potential $V(\sigma)$ is fairly general and the
following discussion 
will not depend on it. The natural choice to keep in mind is a mass term
augmented with the standard
triple and quartic self 
interaction, although one could also consider more complicated logarithmic
potential as is becoming a dilaton field.

Our first observation is 
that the Lagrangian eq.\ref{model} allows for static classical solutions
which are very reminiscent of the Abrikosov - Nielsen - Olesen vortices.
Consider a static field configuration of the form
\begin{eqnarray}
&&G^{ij}=G^{00}=0\nonumber \\
&&G^{0i}=G^{i0}\equiv a^i(\vec x)\nonumber \\
&&\sigma\equiv \rho(\vec x)-v
\end{eqnarray} 
It is easy to see that 
this configuration is a solution of equations of motion, provided $a^i$
and $\rho$ solve 
precisely the same equations as 
in the AHM with the only modification that the
scalar potential is 
given by $V(\sigma)$. We will call this configuration the tensor flux
tube (TFT). In fact it does carry a conserved tensor flux. To see this
let us consider the following operator
\begin{equation}
\tilde F^{\mu\nu\lambda}
=\epsilon^{\mu\nu\rho\sigma}\partial_\rho G_{\sigma\lambda}
\label{F}
\end{equation}
For single valued field $G_{\mu\nu}$ it satisfies the conservation equation
\begin{equation}
\partial_\mu \tilde F^{\mu\nu\lambda}=0
\label{cons}
\end{equation}
This is the analog 
of the homogeneous Maxwell equation of the AHM without monopoles
\begin{equation}
\partial_\mu \tilde F^{\mu\nu}=0
\end{equation}
For our static TFT solution we have
\begin{equation}
\tilde F_{i00}\equiv b_i=\epsilon_{ijk}\partial_j a_k
\end{equation}
and the flux through the plane perpendicular to the symmetry axis of TFT is
\begin{equation}
\Phi\equiv \int dS_i b_i=2\pi/g
\end{equation}
The energy per unit length of the TFT
is directly related to the masses of the two glueballs and to the 
interglueball 
coupling constants $g^2$ and $v$. It is natural to identify this quantity with
the string tension of the pure Yang - Mills theory.

One important qualification 
to the above statements is, that we have assumed that just like in 
the AHM the mass term (and the interaction terms) 
of the tensor field allows quantized discontinuities of the
form $2\pi\Delta$. 
This is not at all unnatural in the present framework. Note that our effective
theory can be thought of as a gauge theory with the ``spontaneously
broken'' gauge group 
of eq.\ref{gauge}. The analog of the phase field $\phi$ of eq.\ref{phi}
in our model is played by 
the gauge parameter $\Lambda_\mu$ of eq.\ref{gauge}. All that is needed
for the discontinuities 
to be allowed is that $\Lambda_\mu$ (or at least $\Lambda_0$) be a 
phase, or in other words 
for the gauge group to have a $U(1)$ subgroup. This does not seem to
be an extremely 
unnatural requirement, although to determine whether this is indeed true
one would need to have 
some information about dynamics on distance scales shorter than the inverse
glueball mass.

Continuing the same 
line of thought, we can identify the objects which should play in our model
the same role as magnetic monopoles in the dual superconductor.
The fields $G$ and $\sigma$ can couple
to much heavier objects, 
which from the low energy point of view are basically pointlike. At these
short distances 
eq.\ref{cons} should be modified, and we can consider the current
\begin{equation}
J^{\nu\lambda}=\partial_\mu \tilde F^{\mu\nu\lambda}
\label{mon}
\end{equation}
Again I stress that 
the current $J^{\nu\lambda}$ must have only very high momentum components
$k>>m$, otherwise eq.\ref{model} 
will not describe faithfully the low energy sector of the theory.
Of course, this 
situation is very similar to AHM with heavy magnetic monopoles where the 
homogeneous Maxwell 
equation is also violated on the distance scales of the order of the monopole
size. 
Now, since $\tilde F^{\mu\nu\lambda}$ 
is antisymmetric under the interchange of $\mu$ and $\nu$,
our newly born current is conserved
\begin{equation}
\partial_\nu J^{\nu\lambda}=0
\end{equation}
 The components of the 
current which are relevant to our TFT configuration are $J^{\mu 0}$.
Imagine that the underlying theory does indeed contain objects that 
carry the charge
$Q=\int d^3x J^{00}$\footnote{At present we do not understand how to 
reconcile the tensorial
nature of the current with the Coleman - Mandula theorem.}. Then the 
argument for confinement 
of these objects is identical to the argument for confinement of magnetic 
monopoles
in AHM. If these objects are identified with heavy quarks this becomes 
the picture of the QCD
confinement from the low energy point of view. Of course, these ``quarks'' 
have very little 
to do with perturbative quarks. But here I appeal to the sentiment from 
the introductory part of 
this paper, namely that we really don't know what should be the relation 
of the perturbative quarks
to the physical confined objects.

This discussion is very far removed from the underlying QCD Lagrangian, 
and for this reason is not
immediatelly helpfull in understanding confinement in terms of the 
ultraviolet degrees of freedom.
This is also not my purpose here. The purpose of this note is to point 
out that the low energy 
effective theory that describes the low lying glueballs contains classical
stringlike objects which are naturally interpreted as the QCD flux tubes. 
To verify this 
interpretation further study is necessary.
I hope, however
that this discussion can teach us two main lessons. First, that the dual 
superconductor
model should not be thought of narrowly as the Abelian Higgs model, but 
rather as a much
larger class of models that admit classical stringlike solutions. 

Second, that the effective
low energy theory of YM theory belongs to this class. In this sense 
our discussion
lends support to the 
extended dual superconductor
hypothesis.

{\bf Acknowledgements.}

I thank Jorge Alfaro and Misha Shifman for discussions.

\end{document}